\documentclass[aps,prd,preprintnumbers,superscriptaddress,nofootinbib,showpacs,twocolumn]{revtex4}%

\usepackage[dvips]{graphicx}
\usepackage{bm,latexsym,amsmath,amssymb,amsfonts,mathrsfs}
\usepackage{color}
\input{colordvi.tex}
\newcommand*{\D}{{\rm d}}

\begin{document}

\title{
Vainshtein screening in a cosmological background in the most general
second-order scalar-tensor theory
}

\author{Rampei~Kimura}
\email[Email: ]{rampei"at"theo.phys.sci.hiroshima-u.ac.jp}
\affiliation{Department of Physical Science, Hiroshima University,
Higashi-Hiroshima 739-8526, Japan
}

\author{Tsutomu~Kobayashi}
\email[Email: ]{tsutomu"at"tap.scphys.kyoto-u.ac.jp}
\affiliation{Hakubi Center, Kyoto University, Kyoto 606-8302, Japan
}
\affiliation{Department of Physics, Kyoto University, Kyoto 606-8502, Japan
}

\author{Kazuhiro~Yamamoto}
\email[Email: ]{kazuhiro"at"hiroshima-u.ac.jp}
\affiliation{Department of Physical Science, Hiroshima University,
Higashi-Hiroshima 739-8526, Japan
}

\begin{abstract}
A generic second-order scalar-tensor theory contains
a nonlinear derivative self-interaction of the scalar degree of freedom $\phi$
{\em\`{a} la} Galileon models, which allows for the Vainshtein screening mechanism.
We investigate this effect on subhorizon scales in a cosmological background,
based on the most general second-order scalar-tensor theory.
Our analysis takes into account all the relevant nonlinear terms
and the effect of metric perturbations consistently.
We derive an explicit form of Newton's constant,
which in general is time-dependent and hence is constrained
from observations, as suggested earlier.
It is argued that in the most general case the inverse-square law
cannot be reproduced on the smallest scales.
Some applications of our results are also presented.
\end{abstract}

\pacs{04.50.Kd, 
95.36.+x
}
\maketitle

\section{Introduction}

Since the discovery of the accelerated expansion of the Universe~\cite{SNe},
numerous attempts have been proposed to explain
the origin of this biggest mystery in modern cosmology.
A vast class of models for cosmic acceleration
invokes a scalar degree of freedom, $\phi$, which may
couple minimally or nonminimally to gravity and ordinary matter.
In the case of nonminimal coupling,
the models are commonly called modified gravity (or ``dark gravity'')
rather than dark energy, as $\phi$ participates
in the long-range gravitational interactions and thereby accelerates
the cosmic expansion.
Modified gravity models must be designed with care,
because otherwise the effect of modification could
persist down to small scales, which could easily be inconsistent
with stringent tests in the solar system and laboratories.
For this reason, screening mechanisms for scalar-mediated force
are crucial.

There are mainly two approaches for screening
the scalar degree of freedom in modified gravity models.
The first one is the Chameleon mechanism~\cite{chameleon},
by which the scalar acquires large mass in a high density environment.
This is employed in viable $f(R)$ gravity~\cite{fr}, which is equivalent to
a scalar-tensor theory with an appropriate potential.
The second one is the Vainshtein mechanism~\cite{vainshtein}. In this case
$\phi$'s kinetic term becomes effectively large in the vicinity of matter
due to some nonlinear derivative interaction,
suppressing the effect of nonminimal coupling.
The Vainshtein screening is typical in Galileon-like models~\cite{gali1} 
and nonlinear massive gravity (e.g., \cite{nlmassive}).
In the Vainshtein case, nonlinearities
play an important role in possible recovery of usual gravity on small scales
even in a weak gravity regime.
To test models of modified gravity against experiments and cosmological observations,
we therefore need to clarify the behavior of gravity
around and below the scale at which the relevant nonlinearities set in.
In this paper, we explore the consequences of the latter mechanism in detail,
taking into account the nonlinear effect.

We study gravity
sourced by a density perturbation of nonrelativistic matter,
on subhorizon scales in a cosmological background,
using the (quasi)static approximation.
To provide generic results, we work in the most general
scalar-tensor theory with second-order field equations~\cite{Horndeski},
which can be derived by generalizing the Galileon theory~\cite{covg, GenGal}
and therefore is expected to be endowed with the Vainshtein mechanism.
In the context of the Galileon,
previous works focus only on the scalar-field equation of motion
to see the profile of (the gradient of) $\phi$,
ignoring gravity backreaction~\cite{gali1, 5thforce}.\footnote{In
the course of the preparation of this manuscript, we became aware of
the very recent paper by De Felice, Kase, and Tsujikawa~\cite{recent},
in which the metric under the influence of the Vainshtein mechanism
is obtained for a static and spherically symmetric
configuration in a subclass of the most general theory.}
Since our approach follows the cosmological perturbation theory
on subhorizon scales~\cite{koyama}, the effect of metric perturbations
can naturally be taken into account consistently.
The results in this paper can be applied to
various aspects of cosmology and astrophysics.

This paper is organized as follows.
We define the theory we consider and then present
the equations governing the background cosmological dynamics
in the next section.
In Sec.~III we derive the perturbation equations
with relevant nonlinear contributions using the subhorizon approximation.
We then explore spherically symmetric solutions of the
perturbation equations in Sec.~IV.
In Sec.~V we present some simple applications of our results 
and finally we conclude in Sec.~VI.
In Appendix A, we summarize the definitions of coefficients 
in the equations in the main text. In Appendix B, we 
discuss a possible variety of solutions of the key equation~(\ref{cubicQ}).
In Appendix C, we present the Fourier transform of the 
perturbation equations.

\section{Cosmology in the most general scalar-tensor theory}

We consider a theory whose action is given by
\begin{eqnarray}
S=\int\D^4x\sqrt{-g}\left({\cal L}_{\rm GG}+{\cal L}_{\rm m}\right),\label{action}
\end{eqnarray}
where
\begin{eqnarray}
{\cal L}_{\rm GG} &=& K(\phi, X)-G_3(\phi, X)\Box\phi
\nonumber\\
&&+G_4(\phi, X)R+G_{4X}\left[(\Box\phi)^2-(\nabla_\mu\nabla_\nu\phi)^2\right]
\nonumber\\
&&+G_5(\phi, X)G_{\mu\nu}\nabla^\mu\nabla^\nu\phi
-\frac{1}{6}G_{5X}\bigl[(\Box\phi)^3
\nonumber\\
&&\qquad\qquad
-3\Box\phi(\nabla_\mu\nabla_\nu\phi)^2+
2(\nabla_\mu\nabla_\nu\phi)^3\bigr],\label{GG}
\end{eqnarray}
with four arbitrary functions,
$K, G_3, G_4,$ and $G_5$, of $\phi$ and $X:=-(\partial\phi)^2/2$.
Here $G_{iX}$ stands for $\partial G_i/\partial X$,
and hereafter we will use such a notation without stating so.
The Lagrangian ${\cal L}_{\rm GG}$ is a mixture of
the gravitational and scalar-field portions, as
the Ricci scalar $R$ and the Einstein tensor $G_{\mu\nu}$
are included.
Note in particular that if a constant piece is present in $G_4$
then it gives rise to the Einstein-Hilbert term.
We assume that matter, described by ${\cal L}_{\rm m}$,
is minimally coupled to gravity.

The Lagrangian~(\ref{GG}) gives the most general scalar-tensor theory
with second-order field equations in four dimensions.
The most general theory was constructed for the first time by
Horndeski~\cite{Horndeski} in a different form than~(\ref{GG}),
and later it was rediscovered by Deffayet {\em et al.}~\cite{GenGal}
as a generalization of the Galileon.
The equivalence of the two expressions is shown by the authors of Ref.~\cite{G2}.
In this paper, we employ the Galileon-like expression~(\ref{GG})
since it is probably more useful than its
original form when discussing the Vainshtein mechanism.
The gravitational and scalar-field equations can be found
in the Appendix of Ref.~\cite{G2}.

We now replicate the cosmological background equations
in the theory~(\ref{action})~\cite{G2, AKT}.
For $\phi=\phi(t)$ and the background metric $\D s^2=-\D t^2+a^2(t)\D\mathbf{x}^2$,
the gravitational field equations are
\begin{eqnarray}
{\cal E}&=&-\rho_{\rm m},\label{frd}
\\
{\cal P}&=&0,
\end{eqnarray}
where
\begin{eqnarray}
{\cal E}&:=&2XK_X-K+6X\dot\phi HG_{3X}-2XG_{3\phi}
\nonumber\\
&&-6H^2G_4+24H^2X(G_{4X}+XG_{4XX})
\nonumber\\&&
-12HX\dot\phi G_{4\phi X}-6H\dot\phi G_{4\phi }
\nonumber\\&&
+2H^3X\dot\phi\left(5G_{5X}+2XG_{5XX}\right)
\nonumber\\&&
-6H^2X\left(3G_{5\phi}+2XG_{5\phi X}\right),
\\
{\cal P}&:=&K-2X\left(G_{3\phi}+\ddot\phi G_{3X} \right)
+2\left(3H^2+2\dot H\right) G_4
\nonumber\\&&
-12 H^2 XG_{4X}-4H\dot X G_{4X}
\nonumber\\&&
-8\dot HXG_{4X}-8HX\dot X G_{4XX}
+2\left(\ddot\phi+2H\dot\phi\right) G_{4\phi}
\nonumber\\&&
+4XG_{4\phi\phi}
+4X\left(\ddot\phi-2H\dot\phi\right) G_{4\phi X}
\nonumber\\&&
-2X\left(2H^3\dot\phi+2H\dot H\dot\phi+3H^2\ddot\phi\right)G_{5X}
\nonumber\\&&
-4H^2X^2\ddot\phi G_{5XX}
+4HX\left(\dot X-HX\right)G_{5\phi X}
\nonumber\\&&
+2\left[2\left(HX\right){\bf \dot{}}+3H^2X\right]G_{5\phi}
+4HX\dot\phi G_{5\phi\phi},
\end{eqnarray}
and
$\rho_{\rm m}$ is the (nonrelativistic) matter energy density,
while the scalar-field equation of motion is
\begin{eqnarray}
{\cal S}:=\dot J+3HJ-P_\phi =0,
\end{eqnarray}
where
\begin{eqnarray}
J&:=&\dot\phi K_X+6HXG_{3X}-2\dot\phi G_{3\phi}
\nonumber\\&&
+6H^2\dot\phi\left(G_{4X}+2XG_{4XX}\right)-12HXG_{4\phi X}
\nonumber\\&&
+2H^3X\left(3G_{5X}+2XG_{5XX}\right)
\nonumber\\&&
-6H^2\dot\phi\left(G_{5\phi}+XG_{5\phi X}\right), 
\\
P_\phi &:=&
K_\phi-2X\left(G_{3\phi\phi}+\ddot\phi G_{3\phi X}\right)
\nonumber\\&&
+6\left(2H^2+\dot H\right)G_{4\phi}
+6H\left(\dot X+2HX\right)G_{4\phi X}
\nonumber\\&&
-6H^2XG_{5\phi\phi}+2H^3X\dot\phi G_{5\phi X}.
\end{eqnarray}
An overdot denotes differentiation with respect to $t$
and $H=\dot a/a$.

\section{Perturbations with relevant nonlinearities}

We work in the Newtonian gauge, in which the perturbed metric is written as
\begin{eqnarray}
\D s^2=-(1+2\Phi)\D t^2+a^2(1-2\Psi)\D \mathbf{x}^2,
\end{eqnarray}
with the perturbed scalar field and matter energy density,
\begin{eqnarray}
\phi&\to&\phi(t)+\delta\phi(t, \mathbf{x}),
\\
\rho_{\rm m}&\to&\rho_{\rm m}(t)[1+\delta(t,\mathbf{x})].
\end{eqnarray}
It will be convenient to use
\begin{eqnarray}
Q:=H\frac{\delta\phi}{\dot\phi},
\end{eqnarray}
which is dimensionless.

We wish to know the behavior of the gravitational and scalar fields
on subhorizon scales sourced by a nonrelativistic matter overdensity $\delta$.
To do so, we may ignore time derivatives in the field equations,
while keeping spatial derivatives. We assume that $\Phi$, $\Psi$, and $Q$
are small, but nevertheless we do not neglect terms that are schematically written as
$(\partial^2\epsilon)^2$ and $(\partial^2\epsilon)^3$,
where $\partial$ represents a spatial derivative and
$\epsilon$ is any of $\Phi, \Psi$, and $Q$.
This is because $L^2(t)\partial^2\epsilon$ could be larger than ${\cal O}(1)$
below certain scales, where $L(t)$ is a typical length scale
associated with $G_i$ which may be as large as the present Hubble radius.
[As we will see, ${\cal O}((\partial^2\epsilon)^4)$ terms do not appear.]


The traceless part of the gravitational field equations
is given by
\begin{eqnarray}
&&{\cal D}_i^{\;j}
\left(
{\cal F}_T\Psi-{\cal G}_T\Phi -
A_1Q\right)
\nonumber\\&&
=
\frac{2B_1}{a^2H^2}Z_i^{\;j}
+\frac{2B_3}{a^2H^2}Y_i^{\;j}
+\frac{2B_3}{a^4H^4}\tilde Z_i^{\;j},
\label{traceless-h}
\end{eqnarray}
where we defined a derivative operator
\begin{eqnarray}
{\cal D}_i^{\;j}:=\partial_i\partial^j-\frac{1}{3}\delta_i^{\;j}\nabla^2,
\quad \nabla^2=\partial_i\partial^i,
\end{eqnarray}
and
\begin{eqnarray}
Z_i^{\;j}&:=&\nabla^2Q{\cal D}_i^{\;j}Q
-\partial_i\partial_kQ\partial^j\partial^kQ+\frac{1}{3}\delta_i^{\;j}
\left(\partial_k\partial_l Q\right)^2,
\nonumber
\\
\\
Y_i^{\;j}&:=&\nabla^2\Phi{\cal D}_i^{\;j}Q
+\nabla^2Q{\cal D}_i^{\;j}\Phi
-\partial_i\partial_kQ \partial^j\partial^k\Phi
\nonumber\\&&
-\partial_i\partial_k\Phi\partial^j\partial^kQ
+\frac{2}{3}\delta_i^{\;j}
\partial_k\partial_l\Phi\partial^k\partial^lQ,
\\
\tilde Z_i^{\;j}&:=&-{\cal Q}^{(2)}{\cal D}_i^{\;j}Q+
2\nabla^2Q\partial_i\partial_kQ\partial^j\partial^kQ
\nonumber\\&&
-2\partial_i\partial_kQ\partial^j\partial_lQ\partial^k\partial^lQ
\nonumber\\&&
+\frac{2}{3}\delta_i^{\;j}\left[
(\partial_k\partial_lQ)^3-\nabla^2Q(\partial_k\partial_l Q)^2
\right].
\end{eqnarray}
The coefficients such as ${\cal F}_T$,
$A_1$, $B_1$, ... that appear in the field equations here and
hereafter are defined in Appendix A.

Applying the operator $\partial_j\partial^i$ to the above quantities, we find
\begin{eqnarray}
\partial_j\partial^i Z_i^{\;j}&=&\frac{1}{6}{\nabla^2}{\cal Q}^{(2)},
\\
\partial_j\partial^i Y_i^{\;j}&=&\frac{1}{3}{\nabla^2}\left(
\nabla^2\Phi\nabla^2Q-\partial_i\partial_j\Phi\partial^i\partial^j Q
\right),
\\
\partial_j\partial^i \tilde Z_i^{\;j}&=&0,
\end{eqnarray}
where
${\cal Q}^{(2)}:=\left(\nabla^2Q\right)^2-\left(\partial_i\partial_j Q\right)^2$.
Thus, from the traceless equation~(\ref{traceless-h}) we obtain
\begin{eqnarray}
&&\nabla^2\left({\cal F}_T\Psi-{\cal G}_T\Phi-A_1 Q\right)
=\frac{B_1}{2a^2H^2}{\cal Q}^{(2)}
\nonumber\\
&&\qquad\qquad
+\frac{B_3}{a^2H^2}\left(
\nabla^2\Phi\nabla^2Q-\partial_i\partial_j\Phi\partial^i\partial^j Q
\right).\label{trlseq}
\end{eqnarray}

The $(00)$ component of the gravitational field equations reads
\begin{eqnarray}
{\cal G}_T\nabla^2\Psi
&=&\frac{a^2}{2}\rho_{\rm m}\delta
-A_2 \nabla^2 Q
-\frac{B_2}{2a^2H^2} {\cal Q}^{(2)}
\nonumber\\&&
-\frac{B_3}{a^2H^2}\left(\nabla^2\Psi\nabla^2Q
-\partial_i\partial_j\Psi\partial_i\partial_jQ\right)
\nonumber\\&&
-\frac{C_1}{3a^4H^4}{\cal Q}^{(3)},\label{00eq}
\end{eqnarray}
where
\begin{eqnarray}
{\cal Q}^{(3)}:=\left(\nabla^2 Q\right)^3
-3\nabla^2Q\left(\partial_i\partial_jQ\right)^2
+2\left(\partial_i\partial_jQ\right)^3.
\end{eqnarray}

Finally, the equation of motion for $\phi$ reduces to
\begin{eqnarray}
&&A_0\nabla^2Q
-A_1\nabla^2\Psi
-A_2\nabla^2\Phi+\frac{B_0}{a^2H^2}{\cal Q}^{(2)}
\nonumber\\&&
-\frac{B_1}{a^2H^2}
\left(\nabla^2\Psi\nabla^2Q-\partial_i\partial_j\Psi\partial^i\partial^jQ\right)
\nonumber\\&&
-\frac{B_2}{a^2H^2}
\left(\nabla^2\Phi\nabla^2Q-\partial_i\partial_j\Phi\partial^i\partial^jQ\right)
\nonumber\\&&
-\frac{B_3}{a^2H^2}
\left(\nabla^2\Phi\nabla^2\Psi -
\partial_i\partial_j\Phi\partial^i\partial^j\Psi \right)
\nonumber\\&&
-\frac{C_0}{a^4H^4}{\cal Q}^{(3)}
-\frac{C_1}{a^4H^4}{\cal U}^{(3)}= 0,\label{seom}
\end{eqnarray}
where
\begin{eqnarray}
{\cal U}^{(3)}&:=&
{\cal Q}^{(2)}\nabla^2\Phi
-2\nabla^2Q\partial_i\partial_jQ\partial^i\partial^j\Phi
\nonumber\\&&
+2\partial_i\partial_jQ\partial^j\partial^kQ\partial_k\partial^i\Phi.
\end{eqnarray}
Equations~(\ref{trlseq}), (\ref{00eq}), and~(\ref{seom}),
supplemented with the matter equations of motion (see Sec.~\ref{mdp})
govern the (quasi)static behavior
of the gravitational potentials and the scalar field
on subhorizon scales.

Note that in deriving Eqs.~(\ref{00eq}) and~(\ref{seom})
we have neglected the ``mass terms''
$(\partial{\cal E}/\partial \phi) \delta\phi$ and
$(\partial{\cal S}/\partial \phi) \delta\phi$.
These contributions could be larger than the higher spatial derivative terms,
and in that case the fluctuation $\delta \phi$ will not be excited.
We do not consider this rather trivial situation
and focus on the case where $\partial{\cal E}/\partial\phi$ and
$\partial{\cal S}/\partial\phi$
can safely be ignored.
Though restricted to the linear analysis,
these terms have been considered in Ref.~\cite{AKT}.

Let us end this section with a short remark.
One may notice that all the terms (except $\delta$) in Eqs.~(\ref{trlseq}),
(\ref{00eq}), and~(\ref{seom}) can be written as total divergences, as
\begin{eqnarray*}
&&\nabla^2\Phi\nabla^2Q-\partial_i\partial_j\Phi\partial^i\partial^jQ
=\partial_i\left(\partial^i\Phi\nabla^2Q-\partial_j\Phi\partial^i\partial^jQ\right),
\\
&&{\cal U}^{(3)}=\partial_i\Bigl(\partial^i\Phi{\cal Q}^{(2)}
-2\partial_j\Phi\nabla^2Q\partial^i\partial^jQ
\nonumber\\&&\qquad\qquad\qquad\qquad\qquad\qquad
+2\partial^j\Phi\partial_{k}\partial_jQ\partial^k\partial^iQ\Bigr).
\end{eqnarray*}
Therefore, those equations can be integrated over a spatial domain ${\cal V}$,
and then one is left with the boundary terms and the enclosed mass,
\begin{eqnarray}
\delta M= \rho_{\rm m}(t)\int_{\cal V}\delta(t,\mathbf{x}')\D^3x',
\end{eqnarray}
as a consequence of neglecting the ``mass terms'' mentioned above.
This fact will be used explicitly in the next section.

\section{Spherically symmetric configurations}

We now want to consider a spherically symmetric overdensity
on a cosmological background.
For this purpose it is convenient to use the coordinate
$r=a(t)\sqrt{\delta_{ij}x^ix^j}$. We are primarily interested in
scales much smaller than the horizon radius, $r H\ll 1$.
Under this circumstance the background metric may be written as
$\D s^2\simeq -\D t^2+\D r^2+r^2\D\Omega^2$, where $\D\Omega^2$ is the
line element of the unit two-sphere.

The spherical symmetry allows us to write
\begin{eqnarray*}
a^{-2}\nabla^2 Q &=&r^{-2}(r^2 Q')',
\\
a^{-4}\left(\nabla^2\Phi\nabla^2Q-\partial_i\partial_j\Phi\partial^i\partial^jQ\right)
&=&2r^{-2}(r\Phi'Q')',
\\
a^{-6}{\cal U}^{(3)}&=&2r^{-2}\left[\Phi'(Q')^2\right]',
\end{eqnarray*}
where a prime denotes differentiation with respect to $r$.
The gravitational field equations and scalar-field equation of motion
can then be integrated once, leading to
\begin{widetext}
\begin{eqnarray}
c_h^2 \frac{\Psi'}{r}-\frac{\Phi'}{r}-\alpha_1\frac{Q'}{r}
&=&\frac{\beta_1}{H^2}
\left(\frac{Q'}{r}\right)^2+2\frac{\beta_3}{H^2} \frac{\Phi'}{r}\frac{Q'}{r},
\label{eq1}\\
\frac{\Psi'}{r}+\alpha_2\frac{Q'}{r}
&=&\frac{1}{8\pi {\cal G}_T}\frac{\delta M(t,r)}{r^3}
-\frac{\beta_2}{H^2}\left(\frac{Q'}{r}\right)^2
-2\frac{\beta_3}{H^2}\frac{\Psi'}{r}\frac{Q'}{r}-\frac{2}{3}
\frac{\gamma_1}{H^4}\left(\frac{Q'}{r}\right)^3,
\label{eq2}\\
\alpha_0\frac{Q'}{r}-\alpha_1\frac{\Psi'}{r}-\alpha_2\frac{\Phi'}{r}
&=&2\left[
-\frac{\beta_0}{H^2}
\left(\frac{Q'}{r}\right)^2+\frac{\beta_1}{H^2}\frac{\Psi'}{r}\frac{Q'}{r}
+\frac{\beta_2}{H^2}\frac{\Phi'}{r}\frac{Q'}{r}+\frac{\beta_3}{H^2}
\frac{\Phi'}{r}\frac{\Psi'}{r}
+\frac{\gamma_0}{H^4}
\left(\frac{Q'}{r}\right)^3+\frac{\gamma_1}{H^4}\frac{\Phi'}{r}\left(\frac{Q'}{r}\right)^2\right],
\nonumber\\
\label{eq3}
\end{eqnarray}
\end{widetext}
where we defined
\begin{eqnarray}
\delta M(t, r)=4\pi\rho_{\rm m}(t)\int^r\delta(t, r)\, {r'}^2\D r',
\end{eqnarray}
$c_h^2:={\cal F}_T/{\cal G}_T$, and dimensionless coefficients
\begin{eqnarray}
\alpha_i(t):=\frac{A_i}{{\cal G}_T},
\quad
\beta_i(t):=\frac{B_i}{{\cal G}_T},
\quad
\gamma_i(t):=\frac{C_i}{{\cal G}_T}.
\end{eqnarray}
Note that $c_h$ is the propagation speed of
gravitational waves which may be different from $1$ in general~\cite{G2}.
Note also that in deriving Eqs.~(\ref{eq1})--(\ref{eq3})
we have set the integration constants to be zero,
requiring that $\Phi'=\Psi'=Q'=0$ is a solution if $\delta M=0$.

For sufficiently large $r$, we may neglect all the nonlinear terms
in the above equations. The solution to the linear equations is given by
\begin{eqnarray}
\Phi'&=&\frac{c_h^2\alpha_0 -\alpha_1^2}{\alpha_0+(2\alpha_1+c_h^2\alpha_2)\alpha_2}
\frac{\mu}{r^2},
\label{linP}
\\
\Psi'&=&\frac{\alpha_0 +\alpha_1\alpha_2}{\alpha_0+(2\alpha_1+c_h^2\alpha_2)\alpha_2}
\frac{\mu}{r^2},
\\
Q'&=&\frac{\alpha_1+c_h^2\alpha_2}{\alpha_0+(2\alpha_1+c_h^2\alpha_2)\alpha_2}
\frac{\mu}{r^2}.
\label{Q-inf}
\end{eqnarray}
where we defined $\mu:=\delta M/8\pi{\cal G}_T$.
In this regime, the parametrized post-Newtonian parameter $\gamma$
is given by
\begin{eqnarray}
\gamma=\frac{\alpha_0+\alpha_1\alpha_2}{c_h^2\alpha_0-\alpha_1^2},
\end{eqnarray}
which in general differs from unity.

\subsection{$G_{4X}=0,\; G_{5}=0$}
\label{G4X0}

A simple example for which nonlinear terms can operate is
the model with $G_{4X}=0=G_5$ and $G_{3X}\neq 0$, {\em i.e.},
\begin{eqnarray}
{\cal L}=G_4(\phi)R+K(\phi, X)-G_3(\phi,X)\Box\phi.\label{gKGB}
\end{eqnarray}
In this case, we have
$\beta_1=\beta_2=\beta_3=\gamma_0=\gamma_1=0$ and
$\quad c_h^2=1$.
We also have the relation
\begin{eqnarray}
\beta_0=\frac{\alpha_1}{2}+\alpha_2\;(\neq 0).\label{relation0}
\end{eqnarray}

The Lagrangian~(\ref{gKGB}) corresponds to
a nonminimally coupled version of kinetic gravity braiding~\cite{KGB},
and has been studied extensively
in the context of inflation~\cite{G-inf}
and dark energy/modified gravity~\cite{GDE, KY1, KY2}.
Of course the nonminimal coupling can be undone
by performing a conformal transformation, but in the present analysis
we have no particular reason to do so.
Note that even in the case of $G_{4}=$const
the scalar $\phi$ is coupled to the curvature at the level of
the field equation, which signals ``braiding.''

Using Eqs.~(\ref{eq1}) and~(\ref{eq2}), Eq.~(\ref{eq3}) reduces to
a quadratic equation
\begin{eqnarray}
\frac{{\cal B}}{H^2}\left(\frac{Q'}{r}\right)^2+\frac{2Q'}{r}
=2{\cal C}\frac{\mu}{r^3},\label{quadratic}
\end{eqnarray}
where
\begin{eqnarray}
{\cal B}:=\frac{4\beta_0}{\alpha_0+2\alpha_1\alpha_2+\alpha_2^2},\quad
{\cal C}:=\frac{\alpha_1+\alpha_2}{\alpha_0+2\alpha_1\alpha_2+\alpha_2^2}.
\end{eqnarray}
Equation~(\ref{quadratic}) can easily be solved to give
\begin{eqnarray}
\frac{Q'}{r}=\frac{H^2}{{\cal B}}\left(
\sqrt{1+\frac{2{\cal B}{\cal C}\mu}{H^2r^3}} -1
\right).
\end{eqnarray}
At short distances, $r^3\ll r_*^3:={\cal B}{\cal C}\mu/H^2$, one finds
\begin{eqnarray}
Q'\simeq \frac{H}{{\cal B}}\sqrt{\frac{2{\cal B}{\cal C}\mu}{r}}.
\end{eqnarray}
In order for this solution to be real, we require
\begin{eqnarray}
{\cal B}{\cal C}>0\;\;\Leftrightarrow\;\;
G_{3X}\left(XG_{3X}+G_{4\phi}\right)>0.
\end{eqnarray}
In this regime, the metric potentials are given by
\begin{eqnarray}
\Phi'&\simeq&\frac{G_N\delta M}{r^2}
-\frac{H(\alpha_1+\alpha_2)}{{\cal B}}
\sqrt{\frac{2{\cal B}{\cal C}G_N\delta M}{r}},
\\
\Psi'&\simeq& \frac{G_N\delta M}{r^2}
-\frac{H\alpha_2}{{\cal B}}\sqrt{\frac{2{\cal B}{\cal C}G_N\delta M}{r}},
\end{eqnarray}
where
\begin{eqnarray}
8\pi G_N:=\frac{1}{2G_4}.
\label{geff_caseA}
\end{eqnarray}
If ${\cal B}{\cal C}\sim {\cal O}(1)$, 
the typical length scale $r_* \sim (\mu/H^2)^{1/3}$ 
can be estimated using Eq.(\ref{geff_caseA}) as
\begin{eqnarray}
  \left(\frac{\mu}{H^2}\right)^{1/3} \simeq 120 
\left(\frac{H_0}{H}\right)^{2/3}
\left(\frac{\delta M}{M_{\odot}}\right)^{1/3}{\rm pc},
\label{vainshtein:estimate}
\end{eqnarray}
where $H_0=70 {\rm km/s/Mpc}$.

Note that in this case the Friedmann equation can be written as
\begin{eqnarray}
3H^2=8\pi G_{\rm cos}\left(\rho_{\rm m}+\rho_\phi\right),
\end{eqnarray}
where $G_{\rm cos}=1/16\pi G_4=G_N$ and
$\rho_\phi=2XK_X-K+6X\dot\phi HG_{3X}-2XG_{3\phi}-6H\dot\phi G_{4\phi}$.
Thus, the effective gravitational coupling governing
short-distance gravity is the same as the one in the Friedmann equation.

The Vainshtein mechanism successfully screens the effect of
the fluctuation $\delta\phi$, so that the two metric potentials
coincide and exhibit the Newtonian behavior at leading order.
However, $G_N$ is time-dependent since
it is a function of the time-dependent field $\phi(t)$,
which means that the Vainshtein mechanism cannot suppress
the time variation of $G_N$ in a cosmological background.
This fact was first noticed in Ref.~\cite{Time-varying}.
We will discuss this point further in the next subsection.

\subsection{$G_{5X}=0$}
\label{G5X0}

Let us consider a class of models with $G_{5X}=0$.
In this case, we see that $\beta_3=\gamma_1=0$.
For the other nonzero coefficients we have the following relations:
\begin{eqnarray}
c_h^2=1+2\beta_1\,(\neq 1),
\quad
\beta_1+\beta_2+2\gamma_0=0.\label{relation1}
\end{eqnarray}

With $G_{5X}=0$,
the problem reduces to solving a cubic equation one can handle.
Indeed, using Eqs.~(\ref{eq1}) and~(\ref{eq2}) to remove $\Phi'$ and $\Psi'$
from Eq.~(\ref{eq3}),
we obtain the cubic equation for $Q'$:
\begin{eqnarray}
&&(Q')^3+{\cal C}_2H^2r(Q')^2+\left(\frac{{\cal C}_1}{2}H^4r^2
-H^2{\cal C}_\beta\frac{\mu}{r}\right)Q'
\nonumber\\&&
-\frac{H^4{\cal C}_\alpha\mu}{2}=0,
\label{cubicQ}
\end{eqnarray}
where $r$-independent coefficients are defined as
\begin{eqnarray}
{\cal C}_\alpha&:=&
\frac{\alpha_1+c_h^2\alpha_2}{2\beta_1\beta_2+c_h^2\beta_2^2-\gamma_0},
\nonumber\\
{\cal C}_\beta&:=&
\frac{\beta_1+c_h^2\beta_2}{2\beta_1\beta_2+c_h^2\beta_2^2-\gamma_0},
\nonumber\\
{\cal C}_1&:=&
\frac{\alpha_0+(2\alpha_1+c_h^2\alpha_2)\alpha_2}{2\beta_1\beta_2+c_h^2\beta_2^2-\gamma_0},
\nonumber\\
{\cal C}_2&:=&
\frac{2\beta_0+3\left(\alpha_1\beta_2+\alpha_2\beta_1+c_h^2
\alpha_2\beta_2\right)}{2\left(2\beta_1\beta_2+c_h^2\beta_2^2-\gamma_0\right)}.
\end{eqnarray}
We note the expressions for $\Psi'$ and $\Phi'$ in terms of $Q'$:
\begin{eqnarray}
\Phi'&=&
c_h^2{\mu \over r^2}
-(\alpha_1+c_h^2\alpha_2)Q'-(\beta_1+c_h^2\beta_2){Q'^2 \over H^2r},\nonumber\\
\Psi'&=&{\mu \over r^2}-\alpha_2Q'-\beta_2{Q'^2 \over H^2r}.
\label{metPert}
\end{eqnarray}

Linearizing Eq.~(\ref{cubicQ}) at $r^3\gg ({\cal C}_\beta/{\cal C}_1)\mu/H^2$,
one obtains the solution~(\ref{Q-inf}) as expected.
This will be matched to one of the following three solutions
at short distances:
\begin{eqnarray}
Q'\simeq
+H\sqrt{{\cal C}_\beta\frac{\mu}{r}},
\quad
-H\sqrt{{\cal C}_\beta\frac{\mu}{r}},
\quad
-\frac{{\cal C}_\alpha}{{\cal C}_\beta}\frac{H^2r}{2}.
\label{solNL}
\end{eqnarray}
If simply
${\cal C}_\alpha\sim{\cal C}_\beta\sim{\cal C}_1\sim{\cal C}_2={\cal O}(1)$,\footnote{This
is probably the most natural case if one considers a model that accounts for
the present cosmic acceleration and the coefficients are evaluated at present time,
because in that case there is only one typical length scale $L=H_0^{-1}$.}
the two regimes are connected at around $r\sim r_*$, where
\begin{eqnarray}
r_*:=\left(\frac{{\cal C}_\alpha^2}{{\cal C}_1^2{\cal C}_\beta}\frac{\mu}{H^2}\right)^{1/3}
\; &&{\rm for~~}Q'\simeq \pm H\sqrt{{\cal C}_\beta\frac{\mu}{r}},
\nonumber\\
r_*:=\left(-\frac{{\cal C}_\beta}{{\cal C}_1}\frac{\mu}{H^2}\right)^{1/3}
\; &&{\rm for~~}Q'\simeq-\frac{{\cal C}_\alpha}{{\cal C}_\beta}\frac{H^2r}{2}.
\label{vainshtein:g5x0}
\end{eqnarray}

\begin{figure}[tb]
  \begin{center}
    \includegraphics[keepaspectratio=true,height=55mm]{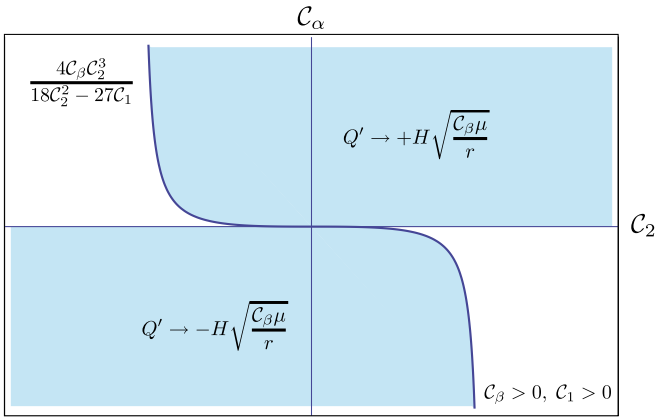}
  \end{center}
  \caption{Relation between the coefficients and the short-distance solution for {\em Case I}.
  In the shaded region one gets a real solution whose behavior
  at short distances is noted.
  }%
  \label{fig:paras1.eps}
\end{figure}

\begin{figure}[t]
  \begin{center}
    \includegraphics[keepaspectratio=true,height=55mm]{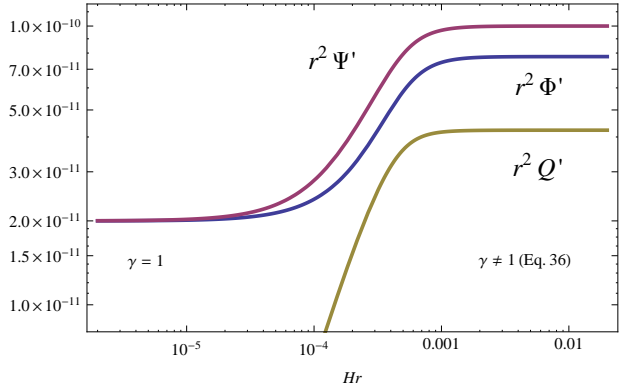}
  \end{center}
  \caption{$r^2\Phi'$, $r^2\Psi'$, and $r^2Q'$ as a function of $r$.
  $H^{-1}=1$. $c_h^2=1.2$, $\alpha_1=\alpha_2=1$, $\alpha_0=2$,
  $\beta_2=2$, $\mu=10^{-10}$. Inside the Vainshtein radius $\gamma=1$
  is reproduced, but the concrete value of the radius depends on the model
  under consideration (see Eqs.~(\ref{vainshtein:estimate}) and (\ref{vainshtein:g5x0})).
  }%
  \label{fig:fig1.eps}
\end{figure}

\begin{figure}[tb]
  \begin{center}
    \includegraphics[keepaspectratio=true,height=55mm]{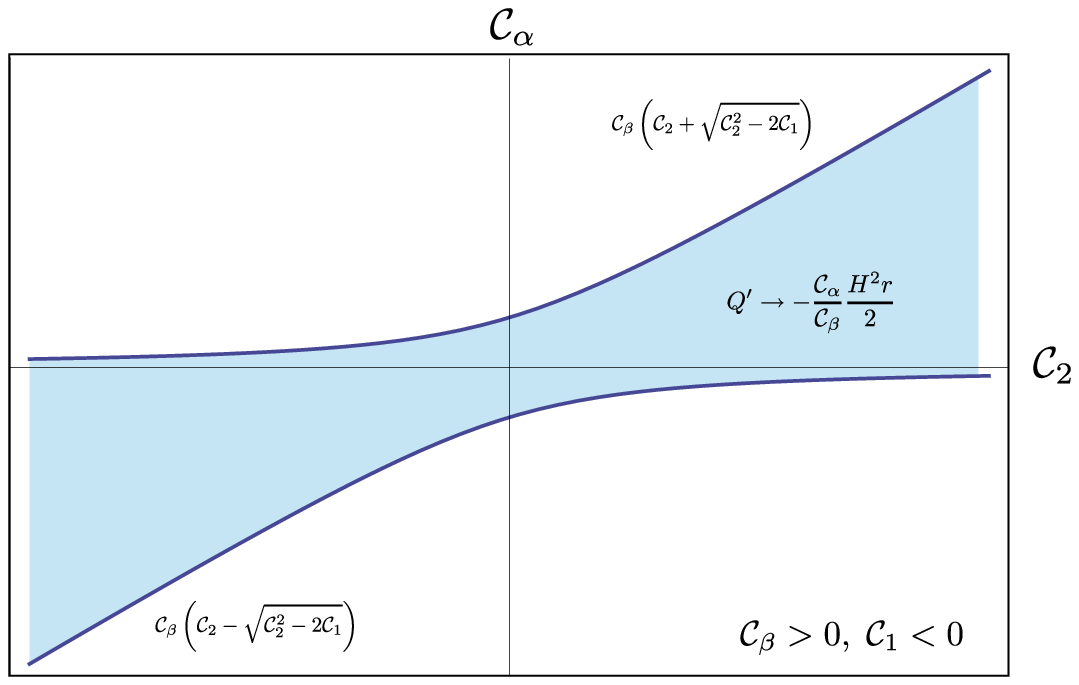}
  \end{center}
  \caption{Coefficients for which a real solution exists ({\em Case II}).
  }%
  \label{fig:paras2.eps}
\end{figure}

\begin{figure}[tb]
  \begin{center}
    \includegraphics[keepaspectratio=true,height=55mm]{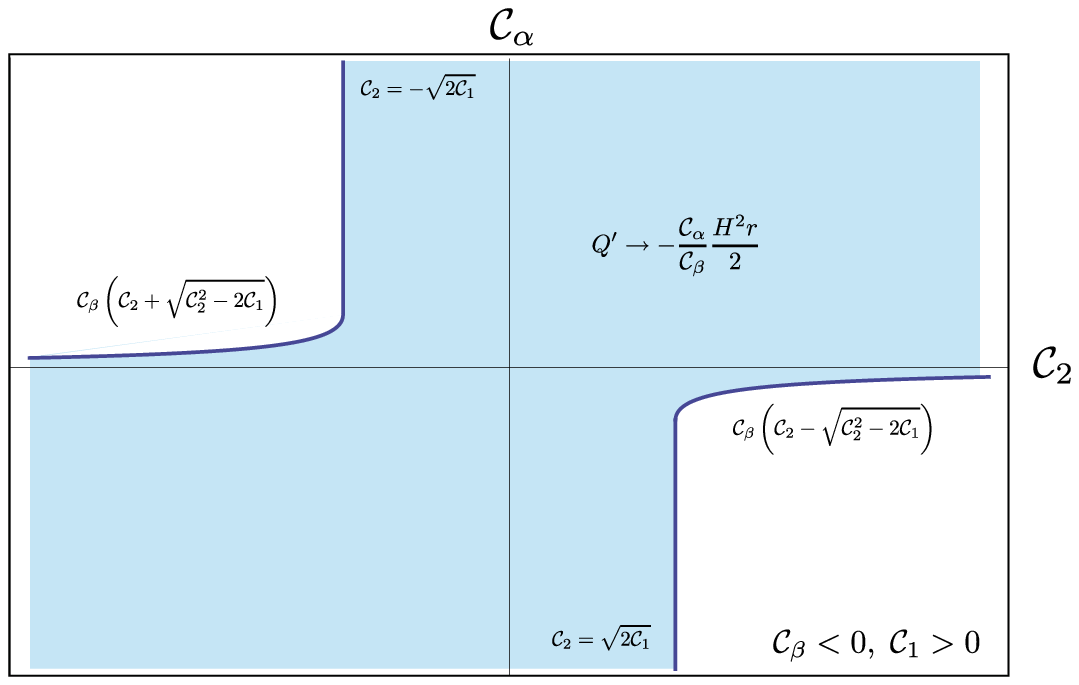}
  \end{center}
  \caption{Coefficients for which a real solution exists ({\em Case II}).
  }%
  \label{fig:paras3.eps}
\end{figure}

If ${\cal C}_\beta>0$ and ${\cal C}_1>0$, the solution with
the boundary condition~(\ref{Q-inf}) at large $r$
can be matched either to $Q'\simeq+H\sqrt{{\cal C}_\beta\mu/r}$
or to $Q'\simeq-H\sqrt{{\cal C}_\beta\mu/r}$. 
We call this situation {\em Case I}.
This is possible for $({\cal C}_2, {\cal C}_\alpha)$
in the shaded region in Fig.~\ref{fig:paras1.eps}.
For ${\cal C}_\alpha>0$ (respectively ${\cal C}_\alpha<0$),
the short-distance solution is given by $+H\sqrt{{\cal C}_\beta\mu/r}$
(respectively $-H\sqrt{{\cal C}_\beta\mu/r}$).
Outside this region one cannot find a solution that is real for $r\in (0, \infty)$.
A typical behavior of the {\em Case I} solution
is plotted in Fig.~\ref{fig:fig1.eps}.
If ${\cal C}_\beta{\cal C}_1<0$, then
the solution can be matched only to $Q'\simeq-({\cal C}_\alpha/{\cal C}_\beta)H^2r/2$.
We call this situation {\em Case II}.
This is possible for $({\cal C}_2, {\cal C}_\alpha)$
in the shaded region in Figs.~\ref{fig:paras2.eps} (${\cal C}_\beta>0,\;{\cal C}_1<0$)
and~\ref{fig:paras3.eps} (${\cal C}_\beta<0,\;{\cal C}_1>0$).
Outside this region no real solutions can be found.
If ${\cal C}_\beta<0$ and ${\cal C}_1<0$ then no real solutions can be found either.



Let us then evaluate the metric perturbations for each solution $Q'$.
We begin with the {\em Case I},
$Q'\simeq\pm\sqrt{{\cal C}_\beta\mu/r}$.
The metric potentials at short distances are given by
\begin{eqnarray}
\Phi'\simeq\frac{C_\Phi}{8\pi {\cal G}_T}\frac{\delta M}{r^2},
\quad
\Psi'\simeq\frac{C_\Psi}{8\pi {\cal G}_T}\frac{\delta M}{r^2},
\end{eqnarray}
where
\begin{eqnarray}
C_\Phi&=&\frac{-\beta_1^2-c_h^2\gamma_0}{2\beta_1\beta_2+c_h^2\beta_2^2-\gamma_0},
\\
C_\Psi&=&\frac{\beta_1\beta_2-\gamma_0}{2\beta_1\beta_2+c_h^2\beta_2^2-\gamma_0}.
\end{eqnarray}
Although the coefficients look apparently different,
now we use the relations~(\ref{relation1}) for the first time
to show that
\begin{eqnarray}
C_\Phi=C_\Psi=\frac{{\cal C}_\beta}{2},
\end{eqnarray}
{\em i.e.,} the two metric potentials actually coincide.
We thus obtain the Newtonian behavior
\begin{eqnarray}
\Phi'\simeq\Psi'\simeq \frac{G_N\,\delta M}{r^2},
\quad
G_N:=\frac{C_\Phi}{8\pi{\cal G}_T}\,(>0).
\label{potential1}
\end{eqnarray}
It is interesting to note that
the above conclusion holds even for the generic propagation speed of
gravitational waves, $c_h^2\neq 1$.
Explicitly, one finds
\begin{eqnarray}
8\pi G_N=\frac{1}{2\left(G_4-4XG_{4X}-4X^2G_{4XX}+3XG_{5\phi}\right)}.
\label{GNewton}
\end{eqnarray}

As in the case of the previous subsection,
$G_N$ is in general time-dependent, as it is a function of
time-dependent $\phi(t)$ and $X=\dot\phi^2(t)/2$.
We thus illustrate how
the Vainshtein mechanism fails to suppress
the time variation of $G_N$ in a cosmological background
within the context of some generic scalar-tensor theories
minimally coupled to matter.
The claim was originally suggested using the Einstein frame
action in Ref.~\cite{Time-varying}.
Here we explicitly give the concrete formula
with which one can evaluate the time variation of $G_N$ for a given model.

If $G_N$ happens to vary very slowly,
we can say that the usual Newtonian gravity is reproduced
in the vicinity of the source.
However, in general, one expects that $G_N$ varies on cosmological time scales.
The time variation $|\dot G/G|$ is constrained from
lunar laser ranging experiments to be
$|\dot G_N/G_N|<0.02 H_0$~\cite{LLR}.

At this stage it is interesting to look at the background evolution
for the models with $G_{5X}=0$.
The (modified) Friedmann equation~(\ref{frd}) can be written as
\begin{eqnarray}
3H^2=8\pi G_{\rm cos}\left(\rho_{m}+\rho_\phi\right),
\end{eqnarray}
where the gravitational coupling in the Friedmann equation read off from the above
exactly coincides with the expression for $G_N$,
\begin{eqnarray}
G_{\rm cos}=G_N,
\end{eqnarray}
and
\begin{eqnarray*}
\rho_\phi&:=&2XK_X-K-2XG_{3\phi}
\nonumber\\&&
+6H
\left(X\dot\phi G_{3X}-2X\dot\phi G_{4\phi X}-\dot\phi G_{4\phi}\right).
\end{eqnarray*}
The situation here is the same as what we have seen in the previous subsection.
We refer to a constraint in Ref.~\cite{uzan}, 
obtained by translating the big bang nucleosynthesis (BBN)
bound on extra relativistic 
degrees of freedom, as
\begin{eqnarray}
\left|1-
\frac{G_N|_{\rm BBN}}{G_N|_{\rm now}}\right|\lesssim 0.1,
\end{eqnarray}
where $G_N|_{\rm BBN}$ (respectively, $G_N|_{\rm now}$)
is evaluated at the time of BBN (respectively, today).

Having thus seen that the Newtonian behavior
is reproduced with time-dependent $G_N$,
let us then evaluate
leading order corrections to the potentials.
In this case we need to keep the subleading term in $Q'$:
\begin{eqnarray}
Q'\simeq\pm H\sqrt{{\cal C}_\beta\frac{\mu}{r}}
+\frac{H^2({\cal C}_\alpha-2{\cal C}_\beta{\cal C}_2)r}{4{\cal C}_\beta}.
\end{eqnarray}
From this we obtain
the corrections $\Delta\Phi'=\Phi'- G_N\delta M/r^2$
and $\Delta\Psi'=\Psi'- G_N\delta M/r^2$ as
\begin{eqnarray}
\Delta\Phi'=
\mp H\left[\alpha_2+\frac{\beta_2}{2{\cal C}_\beta}\left({\cal C}_\alpha-2{\cal C}_\beta{\cal C}_2\right)
\right]\sqrt{\frac{2G_N\delta M}{r}},
\\
\Delta\Psi'=
\mp H\left[\alpha_1+c_h^2\alpha_2+
\frac{\beta_1+c_h^2\beta_2}{2{\cal C}_\beta}\left({\cal C}_\alpha-2{\cal C}_\beta{\cal C}_2\right)
\right]
\nonumber\\
\times
\sqrt{\frac{2G_N\delta M}{r}}.
\end{eqnarray}

For the solution $Q'\simeq -({\cal C}_\alpha/{\cal C}_\beta)H^2r/2$,
we find
\begin{eqnarray}
\Phi'\simeq\frac{c_h^2}{8\pi {\cal G}_T}\frac{\delta M}{r^2}+{\cal O}(r),
\quad
\Psi'\simeq\frac{1}{8\pi {\cal G}_T}\frac{\delta M}{r^2}+{\cal O}(r),
\label{potential2}
\end{eqnarray}
implying that the parametrized post-Newtonian parameter $\gamma$
is given by $\gamma=1/c_h^2$. Therefore, $c_h^2$ is tightly constrained
from solar-system tests in this case: 
$|1-\gamma|<2.3\times 10^{-5}$~\cite{Will}.

When the coefficients ${\cal C}_\alpha,~{\cal C}_\beta,~{\cal C}_1$, and
${\cal C}_2$ have hierarchies in their values, we find a variety
of solutions to Eq.~(\ref{cubicQ}) on an intermediate
scale between the linear regime at large $r$ and the small $r$ limit of
{\em Case I} or {\em Case II}.
The details are summarized in Appendix B, which could be potentially 
confronted with observations.

\subsection{$G_{5X}\neq 0$}

Let us finally discuss the most general case where
all the coefficients in Eqs.~(\ref{eq1})--(\ref{eq3}) are nonzero.
Although one can still eliminate $\Phi'$ and $\Psi'$ to get
an equation solely in terms of $Q'$, the resulting equation
will be a sextic equation.
This hinders us from analyzing a variety of possible solutions in detail.
However, for $\beta_3\neq 0$ and $\gamma_1\neq 0$
one can show that there is no solution such that
$\Phi'\simeq\Psi'\sim 1/r^2$ on sufficiently small scales.
To show this, one substitutes $\Phi'\simeq\Psi'\sim 1/r^2$
to Eq.~(\ref{eq1}). The second term in the right-hand side
can be compensated by the other provided that $Q'\sim r$
or $Q'\sim 1/r^2$.
If $Q'\sim r$, one cannot find a term that compensates
the forth term in the right-hand side of Eq.~(\ref{eq3})
which behaves as $\Phi'\Psi'/r^2\sim 1/r^6$.
If $Q'\sim 1/r^2$, then one cannot find a term that compensates
the last term in the right-hand side of Eq.~(\ref{eq2}).
Thus, there is no consistent solution with
$\Phi'\simeq \Psi'\sim 1/r^2$ on sufficiently small scales.
This implies that the typical length scale associated with $B_3$ and $C_1$
must be as small as ${\cal O}(100\;\mu$m)~\cite{kapner},
though it is uncertain whether or not
we can have the Newtonian behavior of the potentials
on intermediate scales.

\section{Applications}

\subsection{Evolution of density perturbations}\label{mdp}

Since it is assumed that matter is minimally coupled to $\phi$,
no modification is made for the energy conservation equation and
the Euler equation for matter. Therefore, the nonlinear evolution
equation for $\delta$
is given by
\begin{eqnarray}
\ddot\delta+2H\dot\delta-\frac{4}{3}\frac{\dot\delta^2}{1+\delta}
=(1+\delta)\frac{\nabla^2}{a^2}\Phi.
\end{eqnarray}
However, as we have seen in the previous section,
the relation between $\Phi$ and $\delta$ is modified.
One may tackle the nonlinear equations using 
the perturbative approach (e.g., \cite{KTH}). We derive 
the Fourier transform of the nonlinear equations in Appendix C.

For spherical perturbations, $\nabla^2\Phi/a^2$ can be
expressed in terms of $\delta$
using the results in the previous section. It follows that
\begin{eqnarray}
\frac{\nabla^2}{a^2}\Phi \to 
4\pi G_{\rm eff}
\rho_{\rm m}\delta
\end{eqnarray}
where the effective gravitational coupling for large-scale perturbations 
(but well inside the Hubble horizon) is
\begin{eqnarray}
G_{\rm eff} := \frac{1}{8\pi {\cal G}_T}
\frac{c_h^2\alpha_0 -\alpha_1^2}{\alpha_0+(2\alpha_1+c_h^2\alpha_2)\alpha_2},
\end{eqnarray}
and
\begin{eqnarray}
\frac{\nabla^2}{a^2}\Phi
\to 4\pi G_N\rho_{\rm m}\delta
\end{eqnarray}
for small-scale ones.


\subsection{Halo Model}

\begin{figure}[t]
  \begin{center}
    \includegraphics[keepaspectratio=true,height=55mm]{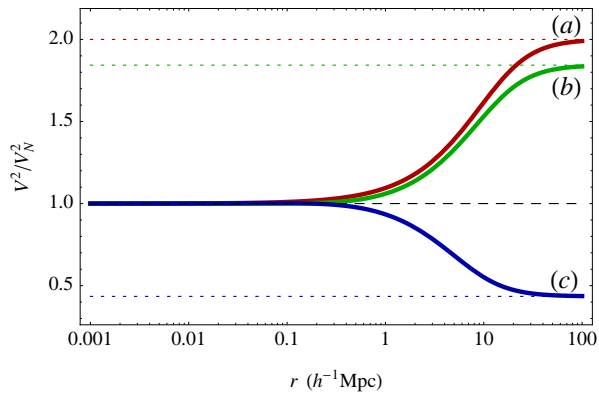}
  \end{center}
  \caption{
The ratio of the circular speed $V^2$ of a test particle in the generalized model 
to that in Newtonian gravity, as a function of radius $r (h^{-1}{\rm Mpc})$.
We used the Hubble parameter at present $H=100 ~h {~\rm km/s/Mpc}$ where $h=0.7$, 
and the velocity dispersion $\sigma=100{~\rm km/s}$. 
Each line corresponds to the following cases: 
$(a)$ $c_h^2=1$, $\alpha_0=-2.0$, $\alpha_2=1.0$,
  $\beta_0=1.0$, $\alpha_1=\beta_1=\beta_2=\gamma_0=0.0$, 
$(b)$ $c_h^2=0.9$, $\alpha_0=-2.0$, $\alpha_1=0.1$, $\alpha_2=0.9$,
  $\beta_0=0.9$, $\beta_1=-0.05$, $\beta_2=0.1$, $\gamma_0=-0.025$,
$(c)$ $c_h^2=0.9$, $\alpha_0=2.0$, $\alpha_1=1.0$, $\alpha_2=0.1$,
  $\beta_0=1.0$, $\beta_1=-0.05$, $\beta_2=0.1$, $\gamma_0=-0.025$.
Dotted and dashed lines are given by Eq.~(\ref{v_L}) and (\ref{v_NL}), 
respectively.
}%
  \label{fig:sishalo.eps}
\end{figure}

We consider a simple halo model to investigate a characteristic
feature of general second-order scalar-tensor theories \cite{Halo}. 
For simplicity, let us assume the density of matter follows
\begin{eqnarray}
\delta\rho(r)={\sigma^2\over 2\pi G_N r^2},
\end{eqnarray}
where $\sigma$ is the parameter of the velocity dispersion. In this 
model we have $\delta M(r)=2\sigma^2r/G_N$. 
This model is the singular isothermal sphere in Newtonian gravity. 

For simplicity, we here consider the cases $G_{4X}=0=G_5$ and $G_{5X}=0$ 
as described in \ref{G4X0} and \ref{G5X0}. 
Although the mass is proportional to $r$, one can check that three 
solutions at short distances remain the same as Eq.~(\ref{solNL}).
To see the effects of modification of gravity, 
we consider the velocity $V(r)$ of a test particle in a circular motion with 
radius $r$, $V^2(r)=r{\Phi'}$,  which reduces to $2\sigma^2$ in 
Newtonian gravity. In the generalized model, $V^2(r)$ depends on 
the radius $r$, whose asymptotic behavior can be found, 
\begin{eqnarray}
V^2\simeq {G_{\rm eff} \over G_N} V_N^2
\label{v_L}
\end{eqnarray}
for large $r$, and 
\begin{eqnarray}
V^2\simeq V_N^2
\label{v_NL}
\end{eqnarray}
for small $r$, where $V_N^2=2\sigma^2$. 
A typical behavior of the circular speed divided by the
one in general relativity, $V^2/V_N^2$, is demonstrated in 
Fig.~{\ref{fig:sishalo.eps}}. The line $(a)$ represents the minimally coupled model,
corresponds to ${G}_4=M_{\rm Pl}^2/2$ and $G_5=0$. 
The lines $(b)$ and $(c)$ show the model with $G_{5X} =0$ and 
the parameters in $(b)$ and $(c)$ are chosen so that the solutions at short distances become
$Q' \simeq -({\cal C}_{\alpha}/{\cal C}_{\beta})H^2r/2$ and $Q'\simeq +H\sqrt{{\cal C}_{\beta}\mu /r}$, respectively.
As one can see in Fig.~{\ref{fig:sishalo.eps}}, the Vainshtein radius $r_* \sim {\cal O}(1)~{\rm Mpc}$ for $\sigma=100$km/s.

\section{Conclusion}

Based on the most general scalar-tensor theory
with second-order field equations, we have studied
metric perturbations on a cosmological background
under the influence of
the Vainshtein screening mechanism.
We have derived the perturbation equations 
with relevant nonlinearities by taking into account the effects of 
cosmological background. 
We have clarified how the Vainshtein mechanism operates in the two subclasses: 
(i) $G_{4X}=G_5=0$ and (ii) $G_{5X}=0$. The situation in the
first case $G_{4X}=G_5=0$ is 
very similar to the Vainshtein mechanism in the Galileon theory,
which contains only $X\square \phi$ in the Lagrangian.
However, the second case $G_{5X}=0$ is considered for the first time
in a cosmological background in this paper.
We have explicitly shown that
below the Vainshtein scale $r_*$ 
there are three possible solutions for $Q'$:
$Q'\simeq \pm H\sqrt{{\cal C}_{\beta}\mu /r}$ and 
$Q' \simeq -({\cal C}_{\alpha}/{\cal C}_{\beta})H^2r/2$.
We have found that two metric perturbations coincide well inside
the Vainshtein radius $r_*$ in the first case,
while the parametrized post-Newtonian parameter $\gamma$ 
is related to the propagation speed of gravitational waves, $c_h^2$,
well inside $r_*$ in the second case.
In both cases, Newton's constant $G_N$, its time variation $|\dot{G}_N/G_N|$, 
and the parametrized post-Newtonian parameter $\gamma$
can be constrained from BBN and the experiments such as lunar laser ranging.
These could provide powerful constraints on the most general scalar-tensor theories.
In the case $G_{5X}\neq 0$,
we have demonstrated that
the inverse-square law cannot be recovered
at sufficiently small $r$.

\acknowledgments 
This work was
supported in part by JSPS Grant-in-Aid for Research Activity Start-up
No.~22840011 (T.K.), the Grant-in-Aid for Scientific Research
No.~21540270 and No.~21244033 and JSPS Core-to-Core Program 
``International Research Network for Dark Energy''.
R.K. acknowledges support by a research assistant program
of Hiroshima University.


\begin{widetext}
\appendix
\section{Coefficients in the field equations}

Here we summarize the definitions of
the coefficients in the field equations:
\begin{eqnarray}
{\cal F}_T&:=&2\left[G_4
-X\left( \ddot\phi G_{5X}+G_{5\phi}\right)\right],
\\
{\cal G}_T&:=&2\left[G_4-2 XG_{4X}
-X\left(H\dot\phi G_{5X} -G_{5\phi}\right)\right],
\\
\Theta&:=&-\dot\phi XG_{3X}+
2HG_4-8HXG_{4X}
-8HX^2G_{4XX}+\dot\phi G_{4\phi}+2X\dot\phi G_{4\phi X}
\nonumber\\&&
-H^2\dot\phi\left(5XG_{5X}+2X^2G_{5XX}\right)
+2HX\left(3G_{5\phi}+2XG_{5\phi X}\right),
\\
A_0&:=&\frac{\dot\Theta}{H^2}+\frac{\Theta}{H}
+{\cal F}_T-2{\cal G}_T-2\frac{\dot{\cal G}_T}{H}-\frac{{\cal E}+{\cal P}}{2H^2},
\\
A_1&:=&\frac{1}{H}\frac{\D{\cal G}_T}{\D t}+ {\cal G}_T-{\cal F}_T,
\\
A_2&:=& {\cal G}_T-\frac{\Theta}{H},
\\
B_0&:=&\frac{X}{H}\biggl\{\dot\phi G_{3X}+3\left(\dot X+2HX\right)G_{4XX}
+2X\dot XG_{4XXX}-3\dot\phi G_{4\phi X}+2\dot\phi XG_{4\phi XX}
\nonumber\\&&
+\left(\dot H+H^2\right)\dot\phi G_{5X}
+\dot\phi
\left[2H\dot X+\left(\dot H+H^2\right) X\right]G_{5XX}
+H\dot\phi X\dot XG_{5XXX}-2\left(\dot X+2HX\right)G_{5\phi X}
\nonumber\\&&
-\dot\phi XG_{5\phi\phi X}-X\left(\dot X-2HX\right)G_{5\phi XX}\biggr\},
\\
B_1&:=&2X\left[G_{4X}+\ddot\phi\left(G_{5X}+XG_{5XX}\right)
-G_{5\phi}+XG_{5\phi X}\right],
\\
B_2&:=&
-2X\left(G_{4X}+2XG_{4XX}+H\dot\phi G_{5X}
+H\dot\phi XG_{5XX}-G_{5\phi}-XG_{5\phi X}\right),
\\
B_3&:=&H\dot\phi XG_{5X},
\\
C_0&:=&2X^2G_{4XX}+\frac{2X^2}{3}\left(2\ddot\phi G_{5XX}
+\ddot\phi XG_{5XXX}-2G_{5\phi X}+XG_{5\phi XX}\right),
\\
C_1&:=&H\dot\phi X\left(G_{5X}+XG_{5XX}\right).
\end{eqnarray}

\section{Intermediate regime for $G_{5X}=0$}

Here we would like to point out that
there could be an interesting intermediate regime
where the quadratic term in Eq.~(\ref{cubicQ}) comes into play
so that we have the solution
\begin{eqnarray}
Q'\simeq \pm H\sqrt{\frac{{\cal C}_\alpha}{2{\cal C}_2}\frac{\mu}{r}}.
\label{interSol}
\end{eqnarray}
This regime can be found in the range
\begin{eqnarray}
\left(\frac{{\cal C}_\beta}{{\cal C}_2^2} \frac{\mu}{H^2} \right)^{1/3}
\ll r\ll \left(\frac{{\cal C}_\alpha{\cal C}_2}{{\cal C}_1^2} \frac{\mu}{H^2} \right)^{1/3}
\; &&{\rm (Case ~I)},
\nonumber\\
\left(\frac{{\cal C}_\beta^2}{{\cal C}_\alpha{\cal C}_2} \frac{\mu}{H^2} \right)^{1/3}
\ll r\ll \left(\frac{{\cal C}_\alpha{\cal C}_2}{{\cal C}_1^2} \frac{\mu}{H^2} \right)^{1/3}
\; &&{\rm (Case ~II)}\label{intermediate}.
\end{eqnarray}
This intermediate regime can be seen if 
${\cal C}_{\alpha}$ or ${\cal C}_{2}$ (${\cal C}_{\beta}$ or ${\cal C}_{1}$) 
is sufficiently large (small) compared with others.
It is also interesting to see the behavior of the
metric perturbations in the intermediate regime~(\ref{intermediate}).
In this regime, the metric potentials can be obtained by substituting 
Eq.~(\ref{interSol}) into the relations~(\ref{metPert}),
\begin{eqnarray}
&&\Phi' \simeq \left[c_h^2-{{\cal C}_{\alpha}\over 2{\cal C}_2}(\beta_1+c_h^2\beta_2)\right]{\mu \over r^2},\\
&&\Psi' \simeq \left(1-\beta_2{{\cal C}_{\alpha}\over 2{\cal C}_2}\right){\mu \over r^2}.
\end{eqnarray}
The metric potentials in this regime differ from Eqs.~(\ref{potential1}) and (\ref{potential2}),
and the parametrized post-Newtonian parameter $\gamma$, 
\begin{eqnarray}
\gamma={2 {\cal C}_2{\cal C}_{\beta}-\beta_2{\cal C}_{\alpha}{\cal C}_{\beta} \over
2 c_h^2 {\cal C}_2{\cal C}_{\beta}-c_h^2 \beta_2{\cal C}_2{\cal C}_{\alpha}-\beta_1{\cal C}_2{\cal C}_{\alpha}},
\end{eqnarray}
is not equal to unity.

We also notice another intermediate regime if ${\cal C}_{\alpha}$ is 
sufficiently large compared with the other coefficients.
In this regime $Q'$ becomes constant,
\begin{eqnarray}
Q'\simeq \left(\frac{H^4 {\cal C}_{\alpha} \mu}{2}\right)^{1/3}.
\label{interSol2}
\end{eqnarray}
This solution can be seen between Eq.~(\ref{interSol}) and Eq.~(\ref{solNL}),
\begin{eqnarray}
\left(\frac{{\cal C}_\beta^3}{{\cal C}_{\alpha}^2} \frac{\mu}{H^2} \right)^{1/3}
\ll r\ll \left(\frac{{\cal C}_\alpha}{{\cal C}_2^3} \frac{\mu}{H^2} \right)^{1/3}.
\label{B7}
\end{eqnarray}
The metric potentials in this regime are given by
\begin{eqnarray}
\Phi' \simeq c_h^2{\mu \over r^2},
\quad
\Psi' \simeq {\mu \over r^2}.
\end{eqnarray}
Thus, the parametrized post-Newtonian parameter is 
$\gamma=1/c_h^2$ in this regime. 
If these regimes (\ref{intermediate}) and (\ref{B7}) include our solar-system scales, 
it is possible to constrain the parametrized post-Newtonian parameter 
$\gamma$ as in the former case.

\section{Equations in the Fourier space}
\def\bfk{{\bf k}}
\def\bfp{{\bf p}}

In this Appendix we summarize the coupled equations for the
evolution of the matter density perturbations in Fourier space, 
which will be useful in the perturbative approach \cite{KTH}.
The matter density perturbations follow
\begin{eqnarray}
&&{\partial \delta \over \partial t} +{1\over a}{\nabla}\cdot[(1+\delta){\bf v}]=0,
\\
&&{\partial {\bf v} \over \partial t} +H {\bf v}
+{1\over a}\left({\bf v}\cdot \nabla\right){\bf v}
=-{1\over a}{\nabla \Phi},
\end{eqnarray}
where ${\bf v}$ is the velocity field. Assuming the irrotational fluid,
we introduce the velocity divergence $\theta=\nabla\cdot {\bf v}/(aH)$. 
Then, due to the Fourier transform, the above equations can be rephrased as
\begin{eqnarray}
&&{1\over H}{\partial \delta(\bfp)\over \partial t}+\theta(\bfp)
=-{1\over (2\pi)^3}\int d\bfk_1 d\bfk_2\delta^{(3)}(\bfk_1+\bfk_2-\bfp)
\left(1+{\bfk_1\cdot\bfk_2\over k_2^2}\right)\theta(\bfk_2)\delta(\bfk_1),
\\
&&{1\over H}{\partial \theta(\bfp)\over \partial t}+
\left(2+{\dot H\over H^2}\right)\theta(\bfp)-{p^2\over a^2H^2}\Phi(\bfp)
\nonumber\\
&&\hspace{4cm}
=-{1\over2}{1\over (2\pi)^3}\int d\bfk_1 d\bfk_2\delta^{(3)}(\bfk_1+\bfk_2-\bfp)
\left({(\bfk_1\cdot \bfk_2)|\bfk_1+\bfk_2|^2\over k_1^2k_2^2}\right)
\theta(\bfk_1)\theta(\bfk_2).
\end{eqnarray}
Similarly, Eqs. (\ref{trlseq}), (\ref{00eq}), and~(\ref{seom}) give
\begin{eqnarray}
&&-p^2\left({\cal F}_T\Psi(\bfp)-{\cal G}_T\Phi(\bfp)-A_1 Q(\bfp)\right)
=\frac{B_1}{2a^2H^2}
{1\over (2\pi)^3}\int d\bfk_1 d\bfk_2\delta^{(3)}(\bfk_1+\bfk_2-\bfp)
\left(k_1^2k_2^2-(\bfk_1\cdot\bfk_2)^2\right)Q(\bfk_1)Q(\bfk_2)
\nonumber\\
&&\qquad\qquad\qquad\qquad
+\frac{B_3}{a^2H^2}
{1\over (2\pi)^3}\int d\bfk_1 d\bfk_2\delta^{(3)}(\bfk_1+\bfk_2-\bfp)
\left(k_1^2k_2^2-(\bfk_1\cdot\bfk_2)^2\right)Q(\bfk_1)\Phi(\bfk_2),
\label{trlseqf}
\end{eqnarray}
\begin{eqnarray}
-p^2{\cal G}_T\Psi(\bfp)
&=&\frac{a^2}{2}\rho_{\rm m}\delta(\bfp)
+p^2A_2 Q(\bfp)
\nonumber\\
&&
-\frac{B_2}{2a^2H^2} 
{1\over (2\pi)^3}\int d\bfk_1 d\bfk_2\delta^{(3)}(\bfk_1+\bfk_2-\bfp)
\left(k_1^2k_2^2-(\bfk_1\cdot\bfk_2)^2\right)Q(\bfk_1)Q(\bfk_2)
\nonumber\\&&
-\frac{B_3}{a^2H^2}
{1\over (2\pi)^3}\int d\bfk_1 d\bfk_2\delta^{(3)}(\bfk_1+\bfk_2-\bfp)
\left(k_1^2k_2^2-(\bfk_1\cdot\bfk_2)^2\right)Q(\bfk_1)\Psi(\bfk_2)
\nonumber\\&&
-\frac{C_1}{3a^4H^4}
{1\over (2\pi)^6}\int d\bfk_1 d\bfk_2d\bfk_3\delta^{(3)}(\bfk_1+\bfk_2+\bfk_3-\bfp)
\nonumber\\&&
\hspace{1cm}\times Q(\bfk_1)Q(\bfk_2)Q(\bfk_3)\biggl[
-k_1^2k_2^2k_3^2+3k_1^2(\bfk_2\cdot\bfk_3)^2-2(\bfk_1\cdot\bfk_2)(\bfk_2\cdot\bfk_3)(\bfk_3\cdot\bfk_1)\biggr],
\label{00eqf}
\end{eqnarray}
\begin{eqnarray}
&&-p^2(A_0Q(\bfp)
-A_1\Psi(\bfp)
-A_2\Phi(\bfp))+\frac{B_0}{a^2H^2}
{1\over (2\pi)^3}\int d\bfk_1 d\bfk_2\delta^{(3)}(\bfk_1+\bfk_2-\bfp)
\left(k_1^2k_2^2-(\bfk_1\cdot\bfk_2)^2\right)Q(\bfk_1)Q(\bfk_2)
\nonumber\\&&
-\frac{B_1}{a^2H^2}
{1\over (2\pi)^3}\int d\bfk_1 d\bfk_2\delta^{(3)}(\bfk_1+\bfk_2-\bfp)
\left(k_1^2k_2^2-(\bfk_1\cdot\bfk_2)^2\right)\Psi(\bfk_1)Q(\bfk_2)
\nonumber\\&&
-\frac{B_2}{a^2H^2}
{1\over (2\pi)^3}\int d\bfk_1 d\bfk_2\delta^{(3)}(\bfk_1+\bfk_2-\bfp)
\left(k_1^2k_2^2-(\bfk_1\cdot\bfk_2)^2\right)\Phi(\bfk_1)Q(\bfk_2)
\nonumber\\&&
-\frac{B_3}{a^2H^2}
{1\over (2\pi)^3}\int d\bfk_1 d\bfk_2\delta^{(3)}(\bfk_1+\bfk_2-\bfp)
\left(k_1^2k_2^2-(\bfk_1\cdot\bfk_2)^2\right)\Phi(\bfk_1)\Psi(\bfk_2)
\nonumber\\&&
-\frac{C_0}{a^4H^4}
{1\over (2\pi)^6}\int d\bfk_1 d\bfk_2d\bfk_3\delta^{(3)}(\bfk_1+\bfk_2+\bfk_3-\bfp)
\nonumber\\&&
\hspace{1cm}\times Q(\bfk_1)Q(\bfk_2)Q(\bfk_3)\biggl[
-k_1^2k_2^2k_3^2+3k_1^2(\bfk_2\cdot\bfk_3)^2-2(\bfk_1\cdot\bfk_2)(\bfk_2\cdot\bfk_3)(\bfk_3\cdot\bfk_1)\biggr]
\nonumber\\&&
-\frac{C_1}{a^4H^4}
{1\over (2\pi)^6}\int d\bfk_1 d\bfk_2d\bfk_3\delta^{(3)}(\bfk_1+\bfk_2+\bfk_3-\bfp)
\nonumber\\&&
\hspace{1cm}\times Q(\bfk_1)Q(\bfk_2)\Phi(\bfk_3)\biggl[
-k_1^2k_2^2k_3^2+(\bfk_1\cdot\bfk_2)^2k_3^2+2k_1^2(\bfk_2\cdot\bfk_3)^2
-2(\bfk_1\cdot\bfk_2)(\bfk_2\cdot\bfk_3)(\bfk_3\cdot\bfk_1)\biggr]
= 0,\label{seomf}
\end{eqnarray}
respectively.

\end{widetext}



\end{document}